%Paper: cond-mat/9509112
%From: janco@stat.th.u-psud.fr (jancovici)
%Date: Tue, 19 Sep 1995 12:26:01 +0100
%Date (revised): Wed, 22 Nov 1995 15:28:47 GMT

\magnification=1200

\expandafter\chardef\csname pre amssym.def at\endcsname=\the\catcode`\@
\catcode`\@=11
\def\hexnumber@#1{\ifcase#1 0\or 1\or 2\or 3\or 4\or 5\or 6\or 7\or 8\or
 9\or A\or B\or C\or D\or E\or F\fi}
\font\tenmsb=msbm10
\font\sevenmsb=msbm7
\font\fivemsb=msbm5
\newfam\msbfam
\textfont\msbfam=\tenmsb
\scriptfont\msbfam=\sevenmsb
\scriptscriptfont\msbfam=\fivemsb
\edef\msbfam@{\hexnumber@\msbfam}
\def\Bbb#1{{\fam\msbfam\relax#1}}
\catcode`\@=\csname pre amssym.def at\endcsname

\input amssym.def
\input amssym
\headline{\ifnum\pageno=1 \nopagenumbers
\else \hss\number \pageno \fi}

\overfullrule=0pt

\font\boldgreek=cmmib10
\textfont9=\boldgreek
\mathchardef\mymu="0916 
\footline={\hfil}
\baselineskip=10pt

\def\R{ {\rm R \kern -.31cm I \kern .15cm}}
\def\C{ {\rm C \kern -.15cm \vrule width.5pt \kern .12cm}}
\def\Z{ {\rm Z \kern -.27cm \angle \kern .02cm}}
\def\N{ {\rm N \kern -.26cm \vrule width.4pt \kern .10cm}}
\def\1{{\rm 1\mskip-4.5mu l} }

\def \cc #1 {\kern .7em \hfill #1 \hfill \kern .7em}
\vbox to 1,5truecm{}
\parskip=0.2truecm
\centerline{\bf The Two-Dimensional Two-Component Plasma Plus}
\medskip
 \centerline{\bf Background on a Sphere : Exact Results}\bigskip

\bigskip
\centerline{{\bf P. J. Forrester}$^1$\footnote{}{$^1$ Department of
Mathematics, University of
Melbourne, Parkville, Victoria 3052, Australia. E-mail~:
matpjf@maths.mu.oz.au}{\bf and B.
Jancovici}$^{1,2}$\footnote{}{$^2$Permanent address : Laboratoire de Physique
Th\'eorique et
Hautes Energies, Universit\'e de Paris XI, b\^atiment 211, 91405 Orsay Cedex,
France (Laboratoire
associ\'e au Centre National de la Recherche Scientifique - URA D0063).
E-mail~:
janco@stat.th.u-psud.fr}}   \medskip

\bigskip\bigskip
\baselineskip=20pt
\noindent
${\bf Abstract}$ \par
An exact solution is given for a two-dimensional model of a Coulomb gas, more
general than
the previously solved ones. The system is made of a uniformly charged
background, positive
particles, and negative particles, on the surface of a sphere. At the special
value $\Gamma
= 2$ of the reduced inverse temperature, the classical equilibrium statistical
mechanics is
worked out~: the correlations and the grand potential are calculated. The
thermodynamic
limit is taken, and as it is approached the grand potential exhibits a
finite-size
correction of the expected universal form.

\vbox to 1cm{}

\vbox to 2cm{}

\noindent LPTHE Orsay 95-48 \par
\noindent June 1995
\vfill\supereject \noindent {\bf 1. INTRODUCTION} \medskip
At the special coupling $\Gamma = q^2/k_BT = 2$, where $q$ denotes the
magnitude of the
charges, $T$ the temperature, and $k_B$ Boltzmann's constant, the classical
two-dimensional
one-component and two-component plasma systems of point particles can be
solved\break
\noindent exactly in a number of different geometries. In the one-component
case$^{(1-3)}$
the calculation is based on the Vandermonde determinant identity and is
performed in the
canonical ensemble, while in the two-component case$^{(4-7)}$ it is based on
the Cauchy
determinant identity and is performed in the grand canonical ensemble. Also, in
the
one-component case the classical Boltzmann factor at $\Gamma = 2$ is
isomorphic$^{(8)}$ to a
squared wave function for $N$ non-interacting fermions in two-dimensions,
subject to a
uniform magnetic field and confined to the lowest Landau level. The grand
partition function
of the two-component plasma at $\Gamma = 2$ is isomorphic$^{(5)}$ to the
generating
functional of the two-dimensional free Fermi field. \par

In Sections 2 and 3 of this paper we unify the exact calculation of the
particle
correlations for one-component$^{(2)}$ and two-component$^{(7)}$ plasma systems
on a
sphere. This is achieved by introducing a new formalism based on a determinant
identity$^{(9)}$ which interpolates between the Vandermonde and Cauchy
identities. In fact
the formalism is of general applicability in the sense that it is not
restricted to the
sphere domain. The\break \noindent exact calculation is performed in an
ensemble with a
uniform background charge density of given total charge $- Nq$, a given number
$N$ of positive
particles, and in addition a variable number $M$ of pairs of positive and
negative particles
which is controlled by a fugacity $\zeta$. In the limit $\zeta \to 0$ the
canonical ensemble
of the one-component plasma is reclaimed, while in the limit $N \to 0$ it is
the grand
canonical ensemble of the two-component plasma which is reclaimed. \par

In the infinite plane system the correlations for the two-component plasma plus
background
system at $\Gamma = 2$ are already known from the work of Cornu and
Jancovici,$^{(10)}$
who used the formalism for the two-component plasma in a field$^{(5)}$ to
perform the
calculation. This suggests an alternative approach, considered in Section 4, to
the
system on a sphere~: at the north pole fix a point charge $Nq$ and then use the
formalism
of ref. 5. Of course the north pole becomes highly singular, as we expect that
$N$
particles of negative charge will accumulate about this point. However, due to
the
screening of the charge-charge correlations, we would expect this not to affect
the
correlations between charges away from the north pole. In fact, we find that
this method
reproduces the exact correlations of Section~3. \par

The grand potential $\Omega_R$ can be calculated by integration with respect to
$\zeta$
of the expression for the density of negative charges. The large-$R$ asymptotic
expansion of $\Omega_R$ is of particular interest as for a Coulomb system on a
sphere
it has been predicted that$^{(6)}$
$$\Omega_R - \Omega_{\infty} \sim {k_BT \over 3} \ln R \ \ \ , \eqno(1.1)$$

\noindent independent of the detail of the plasma. In Section 5, (1.1) is
verified.
\par

In the Appendix, it is shown how the partition function  for the one-component
plasma on a sphere can be calculated within the formalism introduced in Section
2.

\par \bigskip
\noindent {\bf 2. FORMALISM BASED ON A DETERMINANT IDENTITY WHICH\break
INTERPOLATES BETWEEN
THE VANDERMONDE AND CAUCHY\break IDENTITIES} \medskip
In this Section the generalized (with position-dependent fugacities) grand
canonical partition
function for the two-component plasma plus background system at $\Gamma = 2$
will be
expressed as the determinant of an integral operator. This allows the
one-component and
two-component plasma systems to be solved using the same technique. We will
assume that the
domain is the surface of a sphere for calculating the particle correlations in
Section 3.
However, the formalism could also be carried through in the other domains for
which the one-
and two-component plasmas are solvable, in particular the disk and semiperiodic
boundary
conditions. \par

For two charges $q$ and $q'$ on the surface of a sphere of radius $R$,
separated by an
angular distance $\psi$, the pair potential is$^{(2)}$
$$\phi (\psi ) = - qq' \ln \left [ 2R \sin (\psi /2 )/\ell \right ] \ \ \ ,
\eqno(2.1)$$

\noindent where $\ell$ is an arbitrary length scale (to be taken as unity) and
$2R \sin
(\psi /2)$ is equal to the chord length between the particles. \par

Representing the particle positions by spherical polar coordinates $\theta$ and
$\varphi$
($\theta$ is the angle from the north pole and $\varphi$ is usual other polar
angle), and
similarly $\theta '$ and $\varphi '$, we have the identity$^{(2)}$
$$\sin (\psi /2) = |\alpha \beta ' - \alpha ' \beta | = |\beta \beta ' | \left
| {\alpha
\over \beta} - {\alpha ' \over \beta '} \right | \eqno(2.2a)$$

\noindent where
$$\alpha = \cos {\theta \over 2} e^{i \varphi /2} \ \ , \quad \beta = - i \sin
{\theta
\over 2} e^{-i \varphi /2} \ \ \hbox{(and similarly $\alpha '$, $\beta '$)}
\eqno(2.2b)$$

\noindent The coordinates $\alpha$ and $\beta$ are referred to as the
Cayley-Klein
parameters. \par

Consider now a system of $N + M$ particles of charge $q$ with coordinates
specified by
Cayley-Klein parameters $\alpha '_j$, $\beta '_j$ ($j = 1, \cdots , N + M$),
$M$
particles of charge $- q$ with coordinates specified by Cayley-Klein parameters
$\alpha_j$, $\beta_j$ ($j = 1 , \cdots , M$), and a uniform neutralizing
background. From
(2.1) and (2.2), the full Boltzmann factor (including the constants from the
background-particle and background-background interactions) is readily
calculated as
$$W(N + M, N ; \Gamma ) = \left ( {1 \over 2R} \right )^{\Gamma (N/2 + M)} \
e^{\Gamma
N^2/4} \prod_{j=1}^M |\beta_j |^{- \Gamma (1 + N)} \prod_{j=1}^{N+M} |\beta
'_j|^{\Gamma
(N - 1)} |D|^{\Gamma} \eqno(2.3a)$$

\noindent where
$$D : = {\displaystyle{\prod_{1 \leq j < k \leq M}} (u_k - u_j)
\displaystyle{\prod_{1 \leq j
< k \leq N + M}} (v_k - v_j) \over \displaystyle{\prod_{j=1}^M}
\displaystyle{\prod_{k=1}^{N+M}} (u_j - v_k)} \eqno(2.3b)$$

\noindent with
$$u_j := {\alpha_j \over \beta_j} \qquad \hbox{and} \quad v_j := {\alpha '_j
\over \beta
'_j}  \eqno(2.3c)$$

\noindent Thus the generalized grand canonical partition function, with
one-body
potentials with Boltzmann factors $a (\theta ', \varphi ')$ and $b(\theta ,
\varphi )$
coupling to the positive and negative charges respectively, and with $N$ fixed
and $M$
summed over, is given by
$$\Xi_{\Gamma}(a, b) = A_{N\Gamma} \sum_{M=0}^{\infty} {1 \over M !}
{\zeta^{2M} \over (N +
M)!} \left ( {1 \over 2R} \right )^{\Gamma M} R^{4M}$$
$$\times \left ( \prod_{\ell = 1}^{N + M} \int_0^{\pi} d \theta '_{\ell} \sin
\theta
'_{\ell} \left [ \sin (\theta '_{\ell}/2)\right ]^{-\Gamma(1 - N)} \int_0^{2
\pi} d \varphi
'_{\ell} \ a(\theta '_{\ell} , \varphi '_{\ell}) \right )$$
$$\times \left ( \prod_{\ell = 1}^{M} \int_0^{\pi} d \theta_{\ell} \sin
\theta_{\ell}
\left [ \sin (\theta_{\ell}/2) \right ]^{- \Gamma (1 + N)} \int_0^{2 \pi} d
\varphi_{\ell} \ b
(\theta_{\ell} , \varphi_{\ell}) \right ) |D|^{\Gamma} \eqno(2.4a)$$

\noindent where
$$A_{N \Gamma} = R^{2N} \left ( {1 \over 2R} \right )^{N \Gamma / 2} e^{\Gamma
N^2/4} \ \ \ .
\eqno(2.4b)$$

In the case $N = 0$ (two-component plasma) $D$ can be written as a Cauchy
determinant,
while in the case $M = 0$ (one-component plasma) $D$ can be written as a
Vandermonde
determinant. In the general case we have$^{(9)}$
$$D = (-1)^{M(M-1)/2} \det {\Bbb A} \eqno(2.5a)$$

\noindent where
$${\Bbb A} = \left [ \matrix{
1 \hbox to 5.5 truecm{\dotfill} 1 \cr
v_1  \hbox to 4.5 truecm{\dotfill}  v_{M+N}  \cr
\vdots \hskip 5.3 truecm  \vdots  \cr
v_1^{N-1} \hbox to 4 truecm {\dotfill} v_{M+N}^{N-1} \cr
\cr
{1 \over u_1 - v_1} \hbox to 3.5 truecm{\dotfill} {1 \over u_1 - v_{M+N}} \cr
\vdots  \hskip 5 truecm   \vdots \cr
{1 \over u_M - v_1}  \hbox to 3 truecm{\dotfill} {1 \over u_M - v_{M+N}}  \cr
} \right ] \eqno(2.5b)$$

\noindent and consequently
$$|D|^2 = \det \left [ \matrix{
{\Bbb O} &i{\Bbb A} \cr
i{\Bbb A}^+ &{\Bbb O} \cr
} \right ] \eqno(2.6)$$

\noindent where ${\Bbb O}$ denotes the $(M + N) \times (M + N)$ zero matrix.
\par

Consider now the generalized grand partition function (2.4a) for $\Gamma = 2$
with the
substitution (2.6). Suppose we discretize the integrals in (2.4a) by making the
replacements
$$\eqalignno{
&\int_0^{\pi} d \theta_{\ell} \ f(\theta_{\ell} ; \cdots ) \to {1 \over K_1}
\sum_{m_{\ell}=1}^{K_1} f \left ( {\pi (m_{\ell} - \nu_1 ) \over K_1} ; \cdots
\right ) \cr
&\int_0^{\pi} d \theta '_{\ell} \ f(\theta '_{\ell} ; \cdots ) \to {1 \over
K'_1}
\sum_{m'_{\ell}=1}^{K'_1} f \left ( {\pi (m'_{\ell} - \nu '_1 ) \over K'_1} ;
\cdots \right )
\cr &\int_0^{2\pi} d \varphi_{\ell} \ f(\varphi_{\ell} ; \cdots ) \to {1 \over
K_2}
\sum_{n_{\ell}=1}^{K_2} f \left ( {2\pi (n_{\ell} - \nu_2 ) \over K_2} ; \cdots
\right ) \cr
&\int_0^{2\pi} d \varphi '_{\ell} \ f(\varphi '_{\ell} ; \cdots ) \to {1 \over
K'_2}
\sum_{n'_{\ell}=1}^{K'_2} f \left ( {2\pi (n '_{\ell} - \nu '_2 ) \over K'_2} ;
\cdots \right
) &(2.7) \cr  }$$

\noindent where $0 < \nu_1 , \nu '_1, \nu_2 , \nu '_2 < 1$. The domain
available to the
particles now consists of separate sublattices for the positive and negative
charges ($\nu_1,
\cdots , \nu '_2$ are to be chosen so that no two lattice points overlap). We
impose the
constraint that the number of lattice points available to the positive charges,
$K'_1 K'_2$,
is $N$ greater than the number available to the negative charges, $K_1K_2$. The
replacements
(2-7) are exact in the limit $K_1, \cdots , K'_2 \to \infty$ so we can write
\vfill
\supereject
 $$\Xi_2 [a, b] = A_{N2} \ \lim_{K_1, K'_1 \to \infty} \ \lim_{K_2, K'_2 \to
\infty}
\sum_{M=0}^{K_1K_2} {1 \over M!(M + N) !} \left ( {\zeta R \over 2} \right
)^{2M}$$
$$\times \prod_{\ell = 1}^{M + N} \left \{ {1 \over K'_1}
\sum_{m'_{\ell}=1}^{K'_1}
\left [ \sin \pi (m'_{\ell} - \nu '_1)/K'_1 \right ] \left [ \sin \pi
(m'_{\ell} -
\nu '_1)/2K'_1 \right ]^{-2(1 - N)} \right .$$
$$\left . \times {1 \over K'_2} \sum_{n'_{\ell} = 1}^{K'_2} a (\pi (m'_{\ell} -
\nu '_1)/K'_1, 2 \pi (n'_{\ell} - \nu '_2 )/K'_2) \right \}$$
$$\times \prod_{\ell = 1}^{M} \left \{ {1 \over K_1} \sum_{m_{\ell}=1}^{K_1}
\left [ \sin \pi (m_{\ell} - \nu_1)/K_1 \right ] \left [ \sin \pi (m_{\ell} -
\nu_1)/2K_1 \right ]^{-2(1 + N)} \right .$$
$$\left . \times {1 \over K_2} \sum_{n_{\ell} = 1}^{K_2} b (\pi (m_{\ell} -
\nu_1)/K_1, 2
\pi (n_{\ell} - \nu_2 )/K_2) \right \}$$
$$\times \det \left [ \matrix{
{\Bbb O} &i{\Bbb A} \cr
i{\Bbb A}^+ &{\Bbb O} \cr
} \right ] \eqno(2.8)$$

\noindent where the coordinates implicit in ${\Bbb A}$ are taken as lattice
points according
to the prescription (2.7). \par

We now make the crucial observation that expression after the limits in (2.8)
is precisely
the expanded form of a single determinant~:
$$\Xi_2[a, b] = A_{N2} \ \lim_{K_1, K'_1 \to \infty} \ \lim_{K_2, K'_2 \to
\infty} \det \left
( \1 ' + \left [ \matrix{
{\Bbb O} &i{\Bbb B} \cr
i{\Bbb C} &{\Bbb O} \cr
} \right ] \right ) \eqno(2.9)$$

\noindent where $\1 '$ is a $2K'_1K'_2 \times 2K'_1 K'_2$ diagonal matrix with
the first
$N$ diagonal entries zero and the rest one, ${\Bbb O}$ is the $K'_1K'_2 \times
K'_1
K'_2$ zero matrix and
$${\Bbb B} = \left [ \matrix {
\left [ {1 \over K'_1 K'_2} v_{j'k'}^{p-1} \right ]_{p=1, \cdots , N\hfill
\atop{(j'k') =
(11), \cdots , (K'_1 K'_2)}} \hfill \cr
\left [ {(\zeta R/2)b_{jk} \over K'_1 K'_2 (u_{jk} - v_{j'k'})} \right ]_{(jk)
= (11),
\cdots , (K_1 K_2)\hfill \atop{(j'k') = (11), \cdots , (K'_1K'_2)}} \cr }
\right ]
\eqno(2.10a)$$

$${\Bbb C} = \left [ \left [ \bar{v}_{j'k'}^{p-1} \ a_{j'k'} \right ]_{(j'k') =
(11), \cdots , (K'_1 K'_2) \atop{p=1, \cdots , N}\hfill} \left [ {(\zeta R/2)
a_{j'k'} \over
K_1 K_2 (\bar{u}_{jk} - \bar{v}_{j'k'})} \right ]_{(j'k') = (11), \cdots ,
(K'_1K'_2) \atop{(jk) = (11), \cdots , (K_1 K_2)}\hfill} \right ]
\eqno(2.10b)$$

\noindent (the upper subscript on the matrix symbols [$\ $] labels the rows
while the lower
subscript labels the columns), where
$$\eqalignno{
b_{jk} =&\sin \pi (j - \nu_1)/K_1 \left ( \sin \pi (j - \nu_1)/2K_1 \right
)^{-2(1+N)} \cr &\times b \left ( \pi (j - \nu_1)/K_1, 2 \pi (k - \nu_2)/K_2
\right )
&(2.11a) \cr }$$

$$\eqalignno{
a_{j'k'} =&\sin \pi (j' - \nu '_1)/K'_1 \left ( \sin \pi (j' - \nu '_1)/2K'_1
\right
)^{-2(1-N)} \cr &\times a \left ( \pi (j' - \nu '_1)/K'_1, 2 \pi (k' - \nu
'_2)/K'_1 \right
)   &(2.11b) \cr }$$

\noindent and $u_{jk}$ denotes the coordinate $u = \alpha / \beta$ with
$\theta$ and
$\varphi$ therein evaluated at the lattice point $\theta = \pi (j -
\nu_1)/K_1$,
$\varphi = 2 \pi (k - \nu_2)/K_2$ (and similarly for the meaning of $v_{j'k'}$
in terms
of $v = \alpha ' /\beta '$). \par

 From (2.10) and (2.11) we see that the limits in (2.9) can be taken to give
$$\Xi_2[a, b] = A_{N2} \det ({\Bbb X}) \eqno(2.12)$$

\noindent where the operator ${\Bbb X}$ acts on vectors
$$\psi := \left [ \matrix{ a_1 \hfill \cr
\vdots \hfill \cr
a_N \hfill \cr
f(\theta, \varphi ) \hfill \cr
g( \theta , \varphi ) \hfill \cr
}\right ] \eqno(2.13)$$

\noindent and is defined by
$${\Bbb X} \psi := \left [ \matrix{ A_1 \hfill \cr
\vdots \hfill \cr
A_N \hfill \cr
F(\theta, \varphi ) \hfill \cr
G( \theta , \varphi ) \hfill \cr
}\right ] \eqno(2.14)$$

\noindent where
$$A_j := i \int_0^{\pi} d \theta \sin \theta \int_0^{2 \pi} d \varphi \left [
\sin (\theta
/2) \right ]^{N-1} \left [ \cot (\theta / 2) e^{i \varphi}\right ]^{j-1}
g(\theta , \varphi )
\eqno(2.15a)$$ $$\eqalignno{
F(\theta , \varphi ) = &f( \theta , \varphi ) - {i \zeta R \over 2}
\int_0^{\pi}
d\theta_1 \sin \theta_1 \left [ \sin (\theta_1/2) \right ]^{N-1} \left [ \sin
(\theta
/2)\right ]^{-1-N} \cr &\times \int_0^{2 \pi} d \varphi_1 {g( \theta_1,
\varphi_1 ) \
b(\theta_1 , \varphi_1 ) \over \cot (\theta_1/2) \ e^{i\varphi_1 } - \cot
(\theta /2) \
e^{i\varphi}} &(2.15b) \cr }$$

$$\eqalignno{
G(\theta , \varphi ) = &g( \theta , \varphi ) + i \sum_{j=1}^N a_j \left [ \cot
(\theta /2) \ e^{-i \varphi} \right ]^{j-1} [\sin (\theta /2)]^{N-1} a(\theta,
\varphi ) \cr
&+ {i \zeta R \over 2} \int_0^{\pi} d\theta_1 \sin \theta_1 [\sin
(\theta_1/2)]^{-N-1} [\sin
(\theta /2)]^{N-1} \cr  &\times \int_0^{2 \pi} d \varphi_1 {f( \theta_1,
\varphi_1 ) \
a(\theta_1 , \varphi_1 ) \over \cot (\theta_1/2) \ e^{-i\varphi_1 } - \cot
(\theta /2) \
e^{-i\varphi}}   &(2.15c) \cr }$$

In the case $N = 0$, the generalized grand partition function (2.12) reduces to
the known
expression$^{(5)}$ for the grand partition function of the two-component
plasma. In the
other limiting case, $\zeta = 0$, corresponding to the one-component plasma,
the
resulting expression for the generalized partition function is new~; we show in
the
Appendix how to use (2.12) in this case to reclaim the result of ref. 2 for the
exact
evaluation of the partition function. In the general case $N \not= 0$, $\zeta
\not= 0$, we
have not been able to compute the grand potential directly from (2.12), but we
shall obtain it from a density computed in next section. \par \bigskip

\noindent {\bf 3. EVALUATION OF THE CORRELATION FUNCTIONS} \medskip
The fully truncated $n$-particle distribution functions can be obtained by
functional
differentiation of the logarithm of (2.12). Analogous to the situation in the
two-component
plasma limit,$^{(5)}$ the distributions can be expressed in terms of functions
which play
the same role as the Green functions of ref. 5~:
$$G_{s_1s_2} \left ( (\theta_1 , \varphi_1 ), (\theta_2 , \varphi_2 ) \right )
:=
\delta_{s_1s_2} \ \delta \left ( (\theta_1 , \varphi_1), (\theta_2 , \varphi_2)
\right ) -
A_{s_1s_2} \left ( (\theta_1 , \varphi_1 ), (\theta_2 , \varphi_2) \right )
 \eqno(3.1)$$

\noindent where $s_1, s_2 = \pm$, $\delta ((\theta_1, \varphi_1), (\theta_2,
\varphi_2))$
denotes the Dirac delta function on the surface of the sphere~:
$$R^2 \int_0^{\pi} d \theta \sin \theta \int_0^{2 \pi} d \varphi \ H(\theta ,
\varphi ) \
\delta  \left ( (\theta , \varphi ) , (\theta ' , \varphi ') \right ) =
H(\theta ' , \varphi
') \eqno(3.2)$$

\noindent for any continous function $H(\theta , \varphi )$, and $A_{s_1  s_2}$
is
defined in terms of ${\Bbb X}^{-1}$ by
$${\Bbb X}^{-1} = \left [ \matrix{
\alpha_{11}\hbox to 2 truecm{\dotfill} \alpha_{1N}\hfill &F_1 (\theta_2,
\varphi_2
)\hfill &G_1(\theta_2 , \varphi_2)\hfill  \cr
\vdots\hfill \hbox to 1 truecm{} \vdots \hfill &\hskip 0.5 truecm \vdots \hfill
& \hskip 0.5
truecm\vdots \hfill \cr \vdots \hfill \hbox to 1 truecm{} \vdots \hfill&\hskip
0.5 truecm
\vdots \hfill &\hskip 0.5 truecm \vdots  \hfill\cr \vdots \hfill \hbox to 1
truecm{} \vdots
\hfill&\hskip 0.5 truecm \vdots \hfill &\hskip 0.5 truecm \vdots \hfill \cr
\alpha_{N1} \hbox
to 2 truecm{\dotfill} \alpha_{NN}\hfill &F_N(\theta_2 , \varphi_2 ) \hfill
&G_N(\theta_2 ,
\varphi_2 ) \hfill \cr f_1(\theta_1 , \varphi_1 ) \hfill \hbox to 1
truecm{\dotfill}
f_N(\theta_1 , \varphi_1) &A_{++}\left ( (\theta_1 , \varphi_1) , (\theta_2 ,
\varphi_2)
\right ) &A_{+-} \left ( (\theta_1, \varphi_1 ), (\theta_2 , \varphi_2) \right
) \cr
g_1(\theta_1 , \varphi_1 ) \hfill \hbox to 1 truecm{\dotfill} g_N(\theta_1 ,
\varphi_1)
&A_{-+}\left ( (\theta_1 , \varphi_1) , (\theta_2 , \varphi_2) \right ) &A_{--}
\left (
(\theta_1, \varphi_1 ), (\theta_2 , \varphi_2) \right ) \cr
  }\right ] \eqno(3.3)$$

\noindent The one-body density of particles of sign $s$ is
$$\rho_s(\theta , \varphi ) = G_{ss} \left ( (\theta , \varphi ), (\theta,
\varphi ) \right
) \eqno(3.4a)$$

\noindent the two-body truncated density is
$$\rho_{s_1,s_2}^T \left ( (\theta_1 , \varphi_1), (\theta_2 , \varphi_2)
\right ) = -
G_{s_1, s_2} \left ( (\theta_1 , \varphi_1), (\theta_2 , \varphi_2) \right )
G_{s_2, s_1}
\left ( (\theta_2 , \varphi_2 ), (\theta_1, \varphi_1 ) \right )
\eqno(3.4b)$$

\noindent and similarly for the fully truncated $n$-particle distribution. \par

Due to the rotational invariance of the correlations, it suffices to calculate
$\rho_{s_1s_2}^T$ with one of the particles fixed at the south pole
($\rho_{s_1s_2}^T$ will
then depend on $\theta$ only). Thus, from (3.4), it suffices to calculate
$$G_{s_1s_2}\left ( (\theta , \varphi ), (\pi , \varphi ') \right ) :=
G_{s_1s_2}(\theta ,
\varphi ) \ \ \ . \eqno(3.5)$$

 From the definition (3.3) of the $A_{s_1s_2}$ as elements of the inverse of
the operator
${\Bbb X}$ the definition (2.14) of ${\Bbb X}$, and (3.1), it follows that the
functions
$G_{-+}(\theta , \varphi )$ and $G_{++}(\theta , \varphi )$ satisfy the coupled
equations
$$\int_0^{\pi} d \theta \sin \theta \int_0^{2 \pi} d \varphi [\sin (\theta
/2)]^{N-1} [\cot
(\theta /2) \ e^{i \varphi}]^{j-1} G_{++} (\theta , \varphi ) =
\cases{\displaystyle{1 \over
R^2}, \quad j=1 \cr
\cr
0, \qquad j = 2, \cdots , N \cr
} \eqno(3.6a)$$
\vfill \supereject
$$G_{-+}(\theta_1 , \varphi_1 ) + {i \zeta R \over 2} \int_0^{\pi} d \theta_2
\sin \theta_2
[\sin (\theta_2/2)]^{N-1} (\sin \theta_1)^{-1-N}$$
$$\times \int_0^{2 \pi} d \varphi_2 {G_{++} (\theta_2 , \varphi_2 ) \over \cot
(\theta_1 /2) \
e^{i \varphi_1} - \cot (\theta_2 /2) \ e^{i \varphi_2}} = {i \zeta \over 2R}
{[\sin
(\theta_1/2)]^{-1-N} \over \cot (\theta_1/2)} e^{-i \varphi_1} \eqno(3.6b)$$
$$\eqalignno{
&- i  \sum_{j=1}^N \left [ \cot (\theta_1/2) \ e^{-i \varphi_1} \right ]^{j-1}
[\sin
(\theta_1/2)]^{N-1} G_j(\pi ) \cr
&+ {i \zeta R \over 2} \int_0^{\pi} d \theta_2 \sin \theta_2 [\sin
(\theta_2/2)]^{-N-1} [\sin
(\theta_1/2)]^{N-1} \int_0^{2 \pi} {d \varphi_2 \ G_{-+}(\theta_2 , \varphi_2 )
\over \cot
(\theta_2/2) \ e^{-i \varphi_2} - \cot (\theta_1/2) \ e^{-i \varphi_1}} \cr
 &+ G_{++} (\theta_1 , \varphi_1 ) = 0  &(3.6c) \cr }$$

\noindent Also, $G_{+-}(\theta , \varphi )$ and $G_{--}(\theta , \varphi )$
satisfy the same
coupled equations, with the replacements
$$\eqalignno{
&G_{++} (\theta , \varphi ) \to G_{+-}(\theta , \varphi ) \ , \qquad G_{-+}
(\theta
, \varphi ) \to G_{--} (\theta , \varphi ) \cr
&\hbox{r.h.s. of (3.6a) $\to$ 0 each $j = 1 , \cdots , N$, r.h.s. of (3.6b)
$\to$ 0} \cr
&\hbox{r.h.s. of (3.6c) $\to -\displaystyle{i \zeta [\sin (\theta_1/2)]^{N-1}
\over 2R \cot
(\theta_1/2)} e^{i \varphi_1}$}  &(3.7) \cr
}$$

Let us consider the coupled equations (3.6) for $G_{-+}$ and $G_{++}$. We
observe that these
equations permit a solution of the form
$$G_{++}(\theta , \varphi ) = \gamma_{++}(\theta ) \eqno(3.8a)$$
$$G_{-+} (\theta , \varphi ) = \gamma_{-+} (\theta ) \ e^{-i \varphi}
\eqno(3.8b)$$
$$G_j (\pi ) = 0 \ \ \ , \quad j = 2, \cdots , N  \eqno(3.8c)$$

\noindent Substituting (3.8) in (3.6), changing variables $t = \cos \theta$ in
the resulting
equations, and setting
$$\gamma_{-+} (\theta ) = {h_{-+}(t) \over (1 - t)^{N/2} \ (1 + t)^{1/2}}
\eqno(3.9a)$$
$$\gamma_{++} (\theta ) = {h_{++}(t) \over (1 - t)^{(N-1)/2}} \eqno(3.9b)$$
\noindent then gives the coupled equations
$$2^{-(N-1)/2} \int_{-1}^1 dt_2 \ h_{++}(t_2) = {1 \over 2 \pi R^2}
\eqno(3.10a)$$
$$h_{-+}(t) + 2 \pi i \zeta R \int_{-1}^t dt_2 \ h_{++}(t_2) = {i \zeta \over
2R}
\ 2^{(1+N)/2} \eqno(3.10b)$$
$$- i \  2^{-(N-1)/2} G_1(\pi ) + 2 \pi i \zeta R \int_t^1 dt_2 {h_{-+}(t_2)
\over (1
+ t_2) \ (1 - t_2)^N} + (1 - t)^{-(N-1)} h_{++}(t) = 0  \eqno(3.10c)$$

Differentiating (3.10b) and (3.10c) with respect to $t$, substituting the first
of the
resulting equations in the second, and simplifying gives
$$(1 - t^2) \ h''_{-+}(t) + (N-1) (1 + t) \ h'_{-+}(t) - (2 \pi \zeta R)^2 \
h_{-+}(t) = 0
 \eqno(3.11)$$

\noindent This is the Gauss hypergeometric differential equation, in the
variable
$$s = (1 - t)/2  \eqno(3.12)$$

\noindent with (in standard notation)
$$a = - {1 \over 2} \left ( N - (N^2 - 4(2 \pi \zeta R)^2)^{1/2} \right ) \ , \
\  b = -
{1 \over 2} \left ( N + (N^2 - 4(2 \pi \zeta R)^2)^{1/2} \right ) \  , \ \  c =
1 - N \ \
 . \eqno(3.13)$$

\noindent We note from (3.10c) that we require the solution of (3.11) such that
$$h_{-+}(s) = o(s^{N-1}) \qquad \hbox{as} \quad s \to 0 \eqno(3.14)$$

\noindent (otherwise the integral in (3.10c) does not exist~; furthermore
(3.14) ensures
that (3.10a) is satisfied). The solution of (3.11) with this property
is$^{(11)}$
$$h_{-+}(s) = A \ s^N\ F(a + N, b+N ; N+1; s) = A \ s^N(1 - s) \ F(1 - a, 1 -
b; N + 1; s)
\eqno(3.15)$$

\noindent To determine $A$, we note from (3.10b) the requirement
$$h_{-+}(s=1) = {i \zeta \over 2R} \ 2^{1 + {N \over 2}}  \eqno(3.16)$$

\noindent This gives
$$A = {i \zeta \over 2R} \ 2^{1 + {N \over 2}} \ {\Gamma (1 - a) \Gamma (1 - b)
\over \Gamma
(N+1)}  \eqno(3.17)$$

Now that $h_{-+}$ is known, the value of $h_{++}$ follows from differentiating
(3.10b) with
respect to $s$, which gives
$$\eqalignno{
h_{++}(s) &= {1 \over 4 \pi i \zeta R} \ h'_{-+}(s) \cr
&= {1 \over 4 \pi i \zeta R} \  ANs^{N-1} \ F(a+N, b+N ; N; s)  &(3.18) \cr
}$$

Reverting back to the original variable, using (3.12) with $t = \cos \theta$,
and then to
the original Green's function, via (3.9) and (3.8), we thus have
$$G_{-+}(\theta , \varphi ) = e^{-i \varphi} {i \zeta \over 2R} {\Gamma (1 - a)
\Gamma (1 -
b) \over \Gamma (N+1)} \cos (\theta/2) \sin^N (\theta/2) \  F(1 - a, 1 -  b ; N
+ 1 ;
\sin^2 (\theta/2)) \eqno(3.19a)$$

\noindent and, after a simple manipulation on the $\Gamma$ functions,
$$G_{++}(\theta , \varphi ) =  \pi \zeta^2 {\Gamma (N + a) \Gamma (N +
b) \over \Gamma (N)} \sin^{N-1} (\theta/2) \ F(a + N, b+N ; N; \sin^2
(\theta/2))
\eqno(3.19b)$$

To solve the equations (3.6) with the replacements (3.7), we try for a solution
of the form
$$G_{--}(\theta , \varphi ) = \gamma_{--} (\theta ) \eqno(3.20a)$$
$$G_{+-}(\theta , \varphi ) = \gamma_{+-} (\theta ) \ e^{i \varphi}
\eqno(3.20b)$$
$$F_j (\pi ) = 0 \qquad j = 1, \cdots , N  \eqno(3.20c)$$

\noindent Proceeding as in the calculation of $G_{++}$ and $G_{-+}$ above, we
find
$$
G_{+-}(\theta , \varphi ) = \overline{G_{-+}(\theta , \varphi )} \eqno(3.21a)$$

\noindent and
$$\eqalignno{
G_{--}(\theta , \varphi ) = &\pi \zeta^2 {\Gamma (a + N + 1) \Gamma (b +
N + 1) \over \Gamma (N+2)} \sin^{N+1} (\theta/2) \cr
&\times F(a+N+1, b+N+1 ; N+2 ; \sin^2 (\theta/2))  &(3.21b) \cr
}$$

 \par \medskip
\noindent {\bf 4. ALTERNATIVE EVALUATION OF THE CORRELATION FUNCTIONS} \medskip
In this approach, we put on the sphere a uniform background charge density of
total charge
$-Nq$ and a point charge $Nq$ at the north pole, with $N$ fixed. Then, we add a
variable
number $N + M$ of pairs of positive and negative particles. We expect that $N$
 negative particles will stick to the north pole, and therefore there will
remain the
background, $N + M$ positive mobile particles, and $M$ negative mobile
particles, as in the
previous Section. \par

Now, we can use the grand-canonical formalism, with the same variable number $N
+ M$ of
positive and negative particles. The charge $Nq$ at the north pole generates a
one-body
potential, which can be taken into account through a charge and position
dependent fugacity.
{}From (2.1), when $\Gamma := q^2/k_BT = 2$, this fugacity is of the form
$\zeta [\sin^2
(\theta /2)]^{sN}$ for a particle of sign $s (s = \pm 1)$. \par

The problem on the sphere can be transformed into a problem in the plane by the
same
stereographic projection as in ref. 7. Let $P$ be the stereographic projection
of a point
$M$ of the sphere, from the north pole, onto the plane tangent to the south
pole. In terms
of the spherical coordinates $(\theta , \varphi )$ of $M$, the complex
coordinate of $P$ in
the plane is
$$z = 2R \ e^{i \varphi} \cot  {\theta \over 2}  \eqno(4.1)$$

\noindent The projection is a conformal transformation, with a conformal weight
$$e^{\omega} := \sin^2 {\theta \over 2} = {1 \over 1 + {|z|^2 \over 4R^2}}
\eqno(4.2)$$

\noindent That is, an element of length $d\ell$ at $M$ and its projection of
length $|dz|$
at $P$ have a ratio $d\ell/|dz| = e^{\omega}$. Also, in terms of the
coordinates $z$ and
$z'$ of the projections of the particles, the pair potential (2.1) becomes
$$\phi = - qq' \left ( \ln |z - z'| + {\omega \over 2} + {\omega ' \over 2}
\right )
\eqno(4.3)$$

Since each particle interacts with $N + M - 1$ particles of the same sign and
$N + M$
particles of the opposite sign, the $\omega$ term in (4.3) gives a total
one-body potential
$q^2 \omega/2$~; for $\Gamma  = 2$, this generates a factor $e^{-\omega}$ in
the
fugacity. Furthermore, in the computation of the partition function, the area
element $dS$
on the sphere can be expressed in terms of the corresponding area element
$d^2z$ on the
plane as $dS = e^{2 \omega}d^2z$. Altogether, the system on the sphere with a
fugacity
$\zeta e^{\omega sN}$ and the pair potential (2.1) is equivalent to a system in
the plane
with a fugacity
$$\zeta_s (|z|) = \zeta e^{\omega (sN + 1)} = \zeta \left ( 1 + {|z|^2 \over
4R^2} \right
)^{-(sN + 1)} \eqno(4.4)$$

\noindent and a pair potential $- qq' \ln |z - z'|$. The $N$ negative particles
stuck at
the north pole of the sphere are projected at infinity on the plane. \par

The formalism for dealing with a plane system with a fugacity dependent on the
charge and
the position has been set in ref. 5. The $n$-particle densities can be
expressed in terms
of Green functions$^3$\footnote{}{$^3$ In Section 3 of the present paper and in
ref. 5, the Green functions have been defined with different irrelevant phase
factors, which
do not affect the physical quantities.} $g_{s_1s_2}({\bf r}_1, {\bf r}_2)$
(here ${\bf r}_j$
stand for a position vector in the plane~; $z_j = r_j \ e^{i\varphi_j}$)~; in
the plane, the
densities are $2 \pi \zeta e^{\omega (r)} g_{ss}({\bf r}, {\bf r})$, the
two-body truncated
densities are $- s_1s_2 (2 \pi \zeta )^2 e^{\omega(r_1) + \omega(r_2)} |g_{s_1,
s_2}({\bf
r}_1, {\bf r}_2)|^2$. They are the stereographic projections of densities on
the sphere, which
therefore are
$$\rho_s = (2 \pi \zeta ) e^{-\omega (r)} g_{ss}({\bf r}, {\bf r})
\eqno(4.5a)$$
$$\rho_{s_1s_2}^T((\theta_1 , \varphi_1 ), (\theta_2, \varphi_2)) = - s_1s_2 (2
\pi
\zeta)^2 e^{- \omega (r_1) - \omega (r_2)} |g_{s_1s_2}({\bf r}_1, {\bf r}_2)|^2
\eqno(4.5b)$$

\noindent where $(\theta_j , \varphi_j)$ is that point on the sphere which is
projected at
${\bf r}_j$. It is enough to fix one particle at the south pole and to consider
only
$g_{s_1s_2}({\bf r}, 0)$. \par

For the present problem, eq. (2.18a) of ref. 5, with now $m({\bf r}) = 2 \pi
\zeta
e^{\omega (r)}$, becomes
$$\left [ (2 \pi \zeta )^2 \ e^{\omega (r)} + A^+ \ e^{- \omega (r)} A \right ]
g_{++}
({\bf r} , 0) = 2 \pi \zeta \delta ({\bf r}) \eqno(4.6)$$

\noindent where
$$A = e^{i \varphi} \left [  {\partial \over \partial r} + {i \over r}
{\partial \over
\partial \varphi} - {N \over 2} \omega '(r) \right ] \eqno(4.7a)$$

$$A^+ = e^{- i \varphi} \left [ -{\partial \over \partial r} + {i \over r}
{\partial \over
\partial \varphi} - {N \over 2} \omega '(r) \right ]  \eqno(4.7b)$$

\noindent Looking for a solution of the form $g_{++}({\bf r}, 0) = g_{++}(r)$
and changing
to the variable $s = \sin^2 {\theta \over 2} = e^{\omega (r)}$ and to the
function $k(s) =
s^{-N/2} g_{++}$, one obtains from (4.6)
$$s(1 - s) {d^2k \over ds^2} + \left [ N - (N + 1)s \right ] {dk \over ds} - (2
\pi \zeta
R)^2 \ k = 0 \ , \quad s \not=1 \eqno(4.8)$$

\noindent (4.8) is the hypergeometric differential equation. The two solutions
are$^{(11)}$
$F((N+\delta )/2, (N-\delta )/2 ; N ; s)$ and $F((N + \delta )/2, (N - \delta
)/2 ; 1 ;
1 - s)$, where $\delta = (N^2 - 16 \pi^2 \zeta^2 R^2)^{1/2}$. The second
solution behaves
like $s^{-N+1}$ as $s \to 0$~; it would generate in $\rho_{++}^T(\theta , \pi )
= - (2 \pi
\zeta )^2 s^{N-1}|k(s)|^2$ an unacceptable singularity $s^{-N+1}$ as $s =
\sin^2 (\theta /
2) \to 0$, and it must be discarded. Therefore
$$g_{++}({\bf r}, 0) = C_+ s^{N/2} \ F \left ( {N + \delta \over 2} , {N -
\delta \over 2} ; N
; s \right )  \eqno(4.9)$$

\noindent From (4.6), near the south pole $(r = 0, s = 1)$, $g_{++}({\bf r} ,
0)$ behaves
like $- \zeta \ln r \sim -(\zeta /2)\ln (1-~s)$. The $F$ in (4.9) does have a
logarithmic term since
$$F \left ( {N + \delta \over 2}, {N - \delta \over 2} ; N ; s \right ) \sim
{\Gamma (N)
\over \Gamma \left ( {N + \delta \over 2} \right ) \Gamma \left ( {N - \delta
\over 2}
\right )}$$  $$\times \left [ - 2 \gamma - \psi \left ( {N + \delta \over 2}
\right ) - \psi
\left ( {N - \delta \over 2} \right ) - \ln (1 - s) \right ] \ \ , \quad
\hbox{as $s \to 1$}
\eqno(4.10)$$

\noindent where $\Gamma$ is the gamma function, $\psi$ is its logarithmic
derivative, and
$\gamma$ is Euler's constant. The appropriate behaviour is obtained by choosing
$$C_+ = {\Gamma \left ( {N + \delta \over 2} \right ) \Gamma \left ( {N -
\delta \over 2}
\right ) \over \Gamma (N)} {\zeta \over 2}  \eqno(4.11)$$

\noindent Then
$$g_{++}({\bf r}, 0) \sim {\zeta \over 2} \left [ - 2 \gamma - \psi \left ( {N
+ \delta
\over 2} \right ) - \psi \left ( {N - \delta \over 2} \right ) - \ln (1 - s)
\right ] \
\ , \quad \hbox{as $s \to 1$}  \eqno(4.12)$$

By a similar calculation, from eq. (2.18b) of ref. 5, one finds
$$g_{--}({\bf r}, 0) = C_- \ s^{{N \over 2} + 1} F \left ( 1 + {N + \delta
\over 2}, 1 + {N - \delta \over 2} ; N + 2; s \right ) \eqno(4.13)$$

\noindent where
$$C_- = {\Gamma \left ( 1 + {N + \delta
\over 2} \right ) \Gamma \left ( 1 + {N - \delta \over 2} \right ) \over \Gamma
(N + 2)}
{\zeta \over 2} \eqno(4.14)$$

\noindent such that
$$g_{--}({\bf r}, 0) \sim {\zeta \over 2} \left [ - 2 \gamma - \psi \left ( 1 +
{N + \delta
\over 2} \right ) - \psi \left ( 1 + {N - \delta \over 2} \right ) - \ln (1 -
s) \right ] \
\ , \quad \hbox{as $s \to 1$}  \eqno(4.15)$$

\noindent Finally, from eq. (2.18c) of ref. 5, one finds
$$g_{-+}({\bf r}, 0) = {\left ( {N + \delta \over 2} \right ) \left ( {N -
\delta
\over 2} \right ) \over N} {C_+ \over 2 \pi \zeta R} e^{i \varphi} (1 -
s^2)^{1/2}
s^{(N + 1)/2} F \left ( 1 + {N + \delta \over 2} , 1 + {N - \delta \over 2} ; N
+ 1 ;
s \right )  \eqno(4.16)$$

Using the Green functions (4.9), (4.13), (4.16) in (4.5b), one finds
$$\rho_{++}^T (\theta , \pi ) = - \left [ \pi \zeta^2 {\Gamma \left ( {N +
\delta \over 2}
\right ) \Gamma \left ( {N - \delta \over 2} \right ) \over \Gamma (N)}
\sin^{N-1}
{\theta \over 2} F \left ({N + \delta \over 2} , {N - \delta \over 2} ; N ;
\sin^2
{\theta \over 2} \right ) \right ]^2 \eqno(4.17a)$$
$$\rho_{--}^T (\theta , \pi ) = - \left [ \pi \zeta^2 {\Gamma \left ( 1 + {N +
\delta \over 2}
\right ) \Gamma \left ( 1 + {N - \delta \over 2} \right ) \over \Gamma (N+2)}
\sin^{N+1}
{\theta \over 2} \right .$$
$$\left . \times  F \left ( 1 + {N + \delta \over 2} , 1 + {N - \delta \over 2}
; N
+ 2 ; \sin^2 {\theta \over 2} \right ) \right ]^2 \eqno(4.17b)$$

\noindent $\rho_{-+}^T (\theta , \pi ) = \rho_{+-}^T (\theta , \pi )$
\vskip  - 0.5 truecm
$$=\left [ {\zeta \over 2R} {\Gamma \left ( 1 + {N + \delta \over 2}
\right ) \Gamma \left ( 1 + {N - \delta \over 2} \right ) \over \Gamma (N + 1)}
\cos
{\theta \over 2} \sin^{N} {\theta \over 2} F \left ( 1 + {N + \delta \over 2} ,
1 + {N -
\delta \over 2} ; N + 1 ; \sin^2 {\theta \over 2} \right ) \right ]^2
\eqno(4.17c)
$$

\noindent Thus, we retrive the correlation functions derived in Section~3
eqs.~(3.4b), (3.19),
(3.21). \par

It can be easily checked that the limit $\zeta \to 0$ reproduces the
one-component
plasma results of ref. 2 and the limit $N \to 0$ reproduces the two-component
plasma
results of ref.~7. \par

The thermodynamic limit giving a plane system can also be studied~: $N, R \to
\infty$, $\theta \to 0$, with fixed background density $\eta := N/4 \pi R^2$,
fixed
fugacity $\zeta$, and fixed distance $r = 2R \cos (\theta /2)$. By using one of
Kummer's relations$^{(11)}$
$$F(a,b;c;z) = (1 - z)^{-b} \ F \left ( c - a, b ; c; {z \over z - 1} \right )
\eqno(4.18)$$

\noindent one can rewrite (4.17a) as
$$\rho_{++}^T (r) \sim - \left [ n \alpha {\Gamma ( N - \alpha )
\Gamma (\alpha ) \over \Gamma (N)} e^{-{1 \over 2} \pi \eta r^2} N^{\alpha}
(\pi \eta
r^2)^{- \alpha} F \left ( \alpha , \alpha ; N ; 1 -{N \over \pi \eta r^2}
\right ) \right ]^2
\eqno(4.19)$$

\noindent where $\alpha = \pi \zeta^2/\eta$. In the limit $N \to \infty$,
$N^{\alpha}\Gamma (N - \alpha )/\Gamma (N) \to 1$ and
$$(\pi \eta r^2)^{- \alpha} \ F \left ( \alpha , \alpha ; N ; 1 - {N \over \pi
\eta r^2}
\right ) \to e^{{1 \over 2} \pi \eta r^2} (\pi \eta r^2)^{-1/2} \ W_{{1 \over
2}- \alpha ,
0}(\pi \eta r^2) \eqno(4.20)$$

\noindent where $W$ is a Whittaker confluent hypergeometric function.$^{(11)}$
Thus, in
the thermodynamic limit, $$\rho_{++}^T(r) = - \left [  \eta \Gamma (\alpha + 1)
(\pi \eta
r^2)^{-1/2} \ W_{{1 \over 2}- \alpha , 0} (\pi \eta r^2) \right ]^2
\eqno(4.21a)$$

\noindent Similar calculations give
$$\rho_{--}^T(r) = - \left [ \eta \alpha \Gamma (\alpha + 1) (\pi \eta
r^2)^{-1/2} \ W_{-{1
\over 2} - \alpha , 0}(\pi \eta r^2) \right ]^2 \eqno(4.21b)$$
$$\rho_{-+}^T(r) = \rho_{+-}^T = \left [ \eta \alpha^{1/2} \Gamma (\alpha + 1)
(\pi \eta
r^2)^{-1/2} \ W_{- \alpha , {1 \over 2}}(\pi \eta r^2) \right ]^2
\eqno(4.21c)$$

\noindent The results (4.21a) and (4.21b) had been previously
obtained$^4$\footnote{}{$^4$
The sign of the background charge density has been chosen positive in ref. 10,
negative in
the present paper. Furthermore, there are sign inconsistencies at the bottom of
p.~127 of
ref.~10.} in ref.~10.

 \par \medskip \noindent {\bf 5. THERMODYNAMICS} \medskip
In the present generalized grand-canonical formalism, the basic thermodynamic
function is
the generalized grand potential $\Omega$ obtained from the generalized grand
partition
function~:
$$\Omega = - k_B T \ln \Xi  \eqno(5.1)$$

\noindent From (2.4a), the density of negative particles is
$$\rho_- = {<M> \over 4 \pi R^2} = {1 \over 8 \pi R^2} \zeta {d \over d \zeta }
\ln \Xi
\ \ \ . \eqno(5.2)$$

\noindent Thus, we can obtain $\ln \Xi$ and $\Omega$ by integrating
$\rho_-/\zeta$ with
respect to $\zeta$. \par

The densities $\rho_s$, given by (4.5a), should be independent of ${\bf r}$ and
it is
convenient to compute them at ${\bf r} = 0$. However, for point particles,
$g_{ss}(0, 0)$ is
a divergent quantity, as already noticed in simpler cases$^{(5, 7)}$. For
obtaining finite
densities, we assume the particles to be small hard disks of diameter $\sigma$
rather than
point particles. This regularization$^{(5)}$ amounts to replace $g_{ss}(0, 0)$
by
$g_{ss}(\sigma , 0)$. Since $\sigma$ is small, we can use (4.12) and (4.15),
with $1 - s =
\sigma^2/4R^2$, in (4.5a), which gives the densities
$$\rho_+ = \pi \zeta^2 \left [ - 2 \gamma - \psi \left ( {N + \delta \over 2}
\right ) -
\psi \left ( {N - \delta \over 2} \right ) + 2 \ln {2R \over \sigma} \right ]
\eqno(5.3a)$$
$$\rho_- = \pi \zeta^2 \left [ - 2 \gamma - \psi \left ( 1 + {N + \delta \over
2} \right ) -
\psi \left ( 1 + {N - \delta \over 2} \right ) + 2 \ln {2R \over \sigma} \right
]
\eqno(5.3b)$$

\noindent Using $\psi (1 + x) - \psi (x) = 1/x$, one easily checks that $\rho_+
- \rho_-$ is
equal to the background density $\eta = N/4 \pi R^2$. \par

 From now on, we only consider the case of a large system, using large-$R$
expansions for
fixed values of the fugacity $\zeta$ and the background density $\eta = N/4 \pi
R^2$. As
$R$ becomes large, an expansion of the $\psi$ functions gives
$$\rho_- = \rho_+ - \eta = \eta \alpha \left [ - 2 \gamma - \psi (1 + \alpha )
- \ln \pi
\eta \sigma^2 \right ] + {\alpha \over 4 \pi R^2} \left [ {1 \over 2} + \alpha
- \alpha^2
\psi ' (\alpha ) \right ] + O \left ( {1 \over R^4} \right )  \eqno(5.4)$$

\noindent where, again, $\alpha = \pi \zeta^2/\eta$. \par

For $\zeta = 0$, $\Xi$ should become the partition function of the
one-component
plasma$^{(2)}$ (the thermal de Broglie wavelength has been taken as
unity)$^5$\footnote{}{$^5$ Our definition of $Z$ includes the usual $1/N !$
factor which is not present
in the ``excess partition function'' defined in ref. 2.}
$$Z = e^{N^2/2} \ (2 \pi R)^N \ \prod_{p = 0}^{N-1} {p !(N - p - 1)! \over N !}
\eqno(5.5)$$

\noindent as confirmed in the Appendix. The corresponding free energy $\Omega
(0) = -
\beta^{-1} \ln Z$ can be expanded$^{(6)}$ to give
$${1 \over k_BT} \Omega (0) = 4 \pi R^2 {\eta \over 2} \ln {\eta \over 2 \pi^2}
+ {1
\over 3} \ln \left [ (4 \pi \eta )^{1/2} R \right ] + {1 \over 12} - 2 \zeta '
(-1) +
o(1) \eqno(5.6)$$

\noindent where $\zeta '$ is the derivative of Riemann's zeta function. \par

Using (5.4) in (5.2) and integrating from $\zeta = 0$ gives the large-$R$
expansion
$$\eqalignno{
{1 \over k_BT} \Omega = {1 \over k_BT} \Omega (0) &+ 4 \pi R^2 \eta \left [
\alpha (2 \gamma
+ \ln \pi \eta \sigma^2 ) + \ln \Gamma (1 + \alpha ) \right ] \cr
& - {\alpha^2 \over 2} - {\alpha \over 2} + \int_0^{\alpha} x^2 \psi ' (x) dx +
o(1)
&(5.7) \cr
}$$

$(k_BT)^{-1} \Omega$ does exhibit the expected universal finite-size
correction$^{(6)}$ $(1/3) \ln R$. That correction comes entirely from
$\Omega(0)$.

\par \medskip
\noindent {\bf 6. ACKNOWLEDGEMENT} \medskip
We acknowledge the support of the Australian Research Council. Both the French
author (BJ)
and the Australian author (PJF) hope that their collaboration will be pursued,
in spite of
the French nuclear tests in the Pacific, which they jointly disapprove.

\vfill\supereject
\noindent{\bf APPENDIX} \medskip
When $\zeta = 0$ and $a = b = 1$,
$$\Xi_2(a, b) = Z_2 \eqno(A.1)$$

\noindent where $Z_2$ denotes the canonical partition function of the
one-component plasma
on a sphere. Using (2.12) we thus have
$$Z_2 = A_{N2} \prod_k \lambda_k \eqno(A.2)$$

\noindent where the product is over all eigenvalues of the operator ${\Bbb X}$.
The
operator ${\Bbb X}$ again acts on vectors (2.13), except that the component
$f(\theta ,
\varphi )$ is no longer present. The eigenvectors are therefore of the form
$$\psi_k = \left [ \matrix{
a_1^{(k)} \cr
\vdots \cr
a_N^{(k)} \cr
g^{(k)}(\theta , \varphi ) \cr
}\right ] \eqno(A.3)$$

\noindent and from (2.15) the eigenvalues and eigenvectors are specified by the
equations
$$\lambda_k a_j^{(k)} = i \int_0^{\pi} d \theta \sin \theta \int_0^{2 \pi} d
\varphi (\sin
\theta /2)^{N-1} (\cot \theta / 2 \ e^{i \varphi})^{j-1} g^{(k)}(\theta ,
\varphi )
\eqno(A.4)$$

\noindent ($j = 1 , \cdots , N$) and
$$\lambda_k \ g^{(k)}(\theta , \varphi ) = g^{(k)}(\theta , \varphi ) + i
\sum_{j=1}^{N} a_j^{(k)} (\cot \theta /2 \ e^{-i \varphi})^{j-1} (\sin \theta
/2)^{N-1} \ \ \
.\eqno(A.5)$$

We look for solutions of the above equations of the form $$g^{(k)}(\theta ,
\varphi ) = e^{ik \varphi } g_k(\theta ) \ \ \ , \qquad k \in \Z \ \ \ .
\eqno(A.6)$$

\noindent Substituting (A.6) in (A.4) gives
$$a_j^{(k)} = 0 \qquad \qquad \qquad \qquad k \not= 1 - j \eqno(A.7)$$

$$\lambda_k \ a_j^{(k)} = 2 \pi i \int_0^{\pi} d \theta \sin \theta (\sin
\theta /2)^{N-1}
(\cot \theta /2)^{j-1} g_k(\theta ) \ \ , \quad k = 1 - j \ \ \ , \eqno(A.8)$$

\noindent and use of these equations gives from (A.5)
$$(\lambda_k - 1) \ g_k (\theta ) = 0 \ \ , \quad k \notin \{0, \cdots , - (N -
1)\}
\eqno(A.9a)$$

\noindent and
$$\eqalignno{
(1 - \lambda_k) \lambda_k &= 2 \pi \int_0^{\pi} d \theta \sin \theta (\sin
\theta /2)^{2(N-1)}
(\cot \theta /2 )^{-2k} \cr
&= 4 \pi {\Gamma (N + k) \ \Gamma (-k + 1) \over \Gamma (N + 1)} \ \ , \quad k
= 0, - 1 ,
\cdots , - (N-1) \ \ \ . &(A.9b) \cr
}$$

\noindent Eq. (A.9a) gives the eigenvalues $\lambda_k = 1$ for all $k \notin
\{0, \cdots , -
(N-1)\}$, while (A.9b) gives a pair of eigenvalues $\lambda_k^+$ and
$\lambda_k^-$ for each
$k = 0, \cdots , - (N - 1)$ such that their product $\lambda_k^+ \lambda_k^-$
is equal to
the r.h.s. of (A.9b). \par

Thus, writing $p = - k$, we have
$$Z_2 = A_{N2} \prod_{p=0}^{N-1} \left [ 4 \pi {\Gamma (N - p) \ \Gamma (p + 1)
\over \Gamma
(N + 1)} \right ] \eqno(A.10)$$

\noindent which is precisely the result of ref. 2, derived from the Vandermonde
formalism.

\vfill\supereject
\noindent{\bf REFERENCES} \medskip
\item{1.} A. Alastuey and B. Jancovici, {\it J. Phys. (Paris)} {\bf 42} :1
(1981).
\item{2} J. M. Caillol, {\it J. Phys. Lett. (Paris)} {\bf 42} : L-245 (1981).
\item{3.} Ph. Choquard, P. J. Forrester and E. R. Smith, {\it J. Stat. Phys.}
{\bf 33} : 13
(1983).
\item{4.} M. Gaudin, {\it J. Phys. (Paris)} {\bf 46} : 1027 (1985).
\item{5.} F. Cornu and B. Jancovici, {\it J. Chem. Phys.} {\bf 90} : 2444
(1989).
\item{6.} B. Jancovici, G. Manificat and C. Pisani, {\it J. Stat. Phys.} {\bf
76} : 307
(1994).
\item{7.} P. J. Forrester, B. Jancovici and J. Madore, {\it J. Stat. Phys.}
{\bf 69} : 179
(1992).
\item{8.} F. Cornu, B. Jancovici, and L. Blum, {\it J. Stat. Phys.} {\bf 50} :
1221 (1988).
\item{9.} E. L. Basor and P. J. Forrester, {\it Math. Nachr.}  {\bf 170} : 5
(1994).
\item{10.} F. Cornu and B. Jancovici, {\it Europhys. Lett.} {\bf 5} : 125
(1988).
\item{11.} A. Erd\'elyi, {\it Higher Transcendental Functions} (McGraw-Hill,
New York,
1953), vol. 1.
\end